\documentclass[prb,twocolumn,showpacs,showkeys,preprintnumbers,amsmath,amssymb]{revtex4}

\usepackage{amsmath,amssymb,graphicx,multirow,setspace}
\usepackage{color,calc}
\usepackage{footnote}

\begin{document}

\title{Electronic Transport Through EuO Spin Filter Tunnel Junctions}

\author{Nuttachai Jutong$^{1}$}
\author{Ivan Rungger$^{2}$}
\author{Cosima Schuster$^{1}$}
\author{Ulrich Eckern$^{1}$}
\author{Stefano Sanvito$^{2}$}
\author{Udo Schwingenschl\"ogl$^{3}$}
\affiliation{$^{1}$Institut f\"ur Physik, Universit\"at Augsburg, 86135 Augsburg, Germany}
\affiliation{$^{2}$School of Physics and CRANN, Trinity College, Dublin 2, Ireland}
\affiliation{$^{3}$KAUST, PSE Division, Thuwal 23955-6900, Kingdom of Saudi Arabia}

\date{\today}


\begin{abstract}
Epitaxial spin filter tunnel junctions based on the ferromagnetic semiconductor europium monoxide, EuO, are 
investigated by means of density functional theory. In particular, we focus on the spin transport properties of 
Cu(100)/EuO(100)/Cu(100) junctions. The dependence of the transmission coefficient and the current-voltage 
curves on the interface spacing and on the EuO thickness is explained in terms of the EuO density of states and 
the complex band structure. Furthermore we also discuss the relation between the spin transport properties and the 
Cu-EuO interface geometry. The level alignment of the junction is sensitively affected by the interface spacing, 
since this determines the charge transfer between EuO and the Cu electrodes. Our calculations indicate that EuO 
epitaxially grown on Cu can act as a perfect spin filter, with a spin polarization of the current close to 100\%, and 
with both the Eu-5$d$ conduction band and the Eu-4$f$ valence band states contributing to the coherent transport. 
For epitaxial EuO on Cu a symmetry filtering is observed, with the $\Delta_1$ states dominating the transmission. 
This leads to a transport gap larger than the fundamental EuO band gap. Importantly the high spin polarization of 
the current is preserved up to large bias voltages.
\end{abstract}



\maketitle

\section{Introduction}


Magnetic tunnel junctions using MgO as a tunnel barrier generally display large tunneling magneto resistance (TMR) and 
they are now widely used in read heads for ultra high-density hard-disk drives as well as in random access memory 
devices~\cite{S.Yuasa}. Currently a room temperature TMR ratio as high as 500\% can be reached in junctions made
of polycrystalline Fe (more precisely CoFeB) sandwiching a thin MgO insulating layer \cite{S.X.Huang}. Despite such huge 
TMR ratios the Fe/MgO system presents two limiting aspects for future applications. Firstly, there is no flexibility on the
materials side, which means that high TMR ratios are achieved only for the Fe/MgO combination and only for a particular
crystal orientation, (100). Furthermore high crystallinity is needed as the electrodes themselves carry only a moderate 
spin-polarization of the conduction electrons. Secondly, the TMR consistently decreases with increasing the applied 
voltage and/or the temperature and so the operation of a practical device utilizing a magnetic tunnel junction is 
limited to low bias \cite{J.S.Moodera-2007}.

The origin of the high TMR found in Fe/MgO-based junctions is rooted in the spin filtering effect \cite{Butler-2001,Mathon} 
that MgO exerts on the conduction electrons injected from the Fe electrodes
\cite{Ivan-2009,C.Heiliger-2006,Itoh,C.Tiusan-2007,W.H.Butler-2008,G.X.Miao-2009,G.X.Miao-2011}.
Only electrons with specific symmetry and with small transverse momentum contribute significantly to the current. 
For energies around the Fermi energy, $E_F$, these are found only in the majority Fe states. As a 
consequence of this symmetry filtering the decay rate of the conduction electron wave-function into the MgO barrier 
of the majority spins is smaller than that of the minority ones \cite{C.Tiusan-2007}. These decay rates are determined 
by the complex band structure (CBS) of the tunnel barrier \cite{Butler-2001,Mathon}. A crucial feature of the symmetry 
spin filtering effect is that TMR ratios larger than 1000\% can be achieved, despite the fact that the spin-polarization of 
the electrodes is only around 65\% (for CoFeB).

Another possible strategy for obtaining a spin filter effect, and thus very large TMR ratios, is that of using a ferromagnetic 
insulator as spacer between the metallic electrodes. In this case the energy barrier has a different height for majority and 
minority spins, which leads to the suppression of the current for one of the two spin-species for thick enough junctions.
The tunnel current density is proportional to the energy-dependent transmission coefficient through the barrier, $T(E)$, 
which itself depends exponentially on the barrier height, $\Phi$, and the barrier thickness, $t$. This can be written as
$T(E)\propto \exp[-2 \kappa(\Phi,E) t]$, with $\kappa(\Phi, E)=\sqrt{2m_e(\Phi-E)}/\hbar$, with $m_e$ being the electron 
mass. Clearly if the barrier height is different for the different spin-species the current polarizations will increase
exponentially with the insulating layer thickness, leading to full spin polarization for thick barriers \cite{J.S.Moodera-2007,G.X.Miao-2011}. 
Devices constructed with this principles are called spin filter tunnel junctions (SFTJs). The quest for manufacturing SFTJs
then reduces to that of finding suitable ferromagnetic semiconductors.

The europium chalcogenides EuS, EuSe and EuO have all a rocksalt structure and they are all ferromagnetic
insulators. Among these EuO presents the largest conduction band exchange splitting, $\sim$0.54~eV, below the 
material's Curie temperature of 69~K \cite{A. Mauger,Steeneken-2002,J.S.Moodera-2004}. In EuO the divalent Eu ions 
possess a large local moment originating from the half-filled 4$f$ band ($\mu_{\rm Eu}=7\mu_B$). An energy gap of 
1.1~eV separates the half-filled majority Eu-4$f$ band from the Eu-5$d$ conduction band \cite{J.S.Moodera-2007}. 
SFTJs based on polycrystalline EuO in the form of a metal/EuO/metal heterojunction have been studied in
several recent experiments \cite{J.S.Moodera-2004,E.Negusse-J.Appl-2006,SANTOS-2008,Watson-2008,E.Negusse-J.Appl-2009,Martina-J.Appl-2009,Martina-Europhys-2009}. However, the spin transport properties of crystalline epitaxial EuO junctions have
not been studied theoretically so far. The purpose of this paper is to provide such theoretical insight. In particular we present \textit{ab-initio} results for the electronic structure and the electron transmission 
through EuO barriers sandwiched between Cu electrodes oriented along the [001] direction.

Our paper is organized as follows. We start our discussion by presenting the methods used and the structure of the 
device investigated. Then, in the following section we present the electronic structure of EuO and its complex band 
structure along the [001] direction. This determines the spin-dependent decay rates and thus the spin filter efficiency.
In section IV we discuss the transmission coefficient at zero bias, while in section V the dependence of the spin transport 
on the EuO thickness is analyzed and related to the complex band structure. The current-voltage, $I$-$V$, characteristics
are discussed in the following section, where we find that, in contrast to Fe/MgO junctions, the polarization of the current is 
not reduced by the applied bias. Finally we conclude.

\section{Methods}
\label{sec:methods}

The EuO and Cu electronic structures are calculated by using the \textit{ab-initio} density functional theory (DFT) code 
SIESTA \cite{siesta}. Since the local density approximation (LDA) is not sufficient to give an accurate description of the 
EuO density of states (DOS), we correct for on-site Coulomb repulsion with an LDA+U treatment \cite{anis91,anismov93}. 
Following Ref.~[\onlinecite{ingle08}], the exchange constant $J$ and the on-site orbital potential $U$ for the Eu-4$f$ orbitals 
are set respectively to $J_f=0.77$~eV and $U_f=8.3$~eV, while for the O-2$p$ orbitals we use $J_p=1.2$~eV and 
$U_p=4.6$~eV. Troullier-Martins norm-conversing relativistic pseudopotentials are used for Cu, Eu and O. The wave functions 
are expanded over a double $\zeta$ plus polarization basis set (DZP), except for the Eu-4$f$ states, for which we use single-$\zeta$. 
An equivalent real space mesh cutoff of 600~Ry is used together with an electronic temperature of 69~K. We sample the 
Brillouin zone (BZ) in the plane perpendicular to the transport direction over a $7\times7$ $k$-point uniform mesh.

Spin transport is studied by using the SMEAGOL electronic transport code \cite{Smeagol1,Rocha,Ivan}, which combines DFT with the 
nonequilibrium Green's functions (NEGF) technique. SMEAGOL uses the Hamiltonian matrix provided by SIESTA to calculate the charge 
density so that the same pseudopotentials and exchange and correlation function can be used for both the electronic structure and
the transport. The spin current at each bias voltage is calculated by energy integration of the spin-dependent transmission coefficient 
$T^{\sigma}$,
\begin{equation}
I^{\sigma}=\frac{e}{h} {\int{dE\;T^{\sigma}(E,V)\left[f\left(E+\frac{eV}{2}\right)-f\left(E-\frac{eV}{2}\right)\right]}}\:,
\end{equation}
where $\sigma$ is the spin index ($\sigma=\uparrow, \downarrow$), $f$ is the Fermi function, $V$ is the applied bias voltage and 
$e$ the electron charge. The total transmission coefficient is obtained by integrating the $k$-dependent transmission 
$T^{\sigma}(E,V,k_x,k_y)$ over the 2-dimensional BZ perpendicular to the transport direction,
\begin{equation}
T^{\sigma}(E,V)=\frac{1}{\Omega}\int{dk_{x}dk_y\;T^{\sigma}(E,V,k_x,k_y)}\:,
\end{equation}
where $\Omega$ is the area of the BZ. We denote $T(E)=T(E,V=0)$ as the zero bias transmission coefficient.
In our calculations we assume the lattice structure to be periodic in the $x$-$y$ plane, and we keep the transport direction 
along the $z$-axis. While $7\times7$ $k$-points are enough to accurate converge the charge density, the presence of resonances 
in $T^{\sigma}(E,V,k_x,k_y)$ requires the much larger mesh of $100\times100$ to evaluate the transmission coefficient both 
for zero and finite bias. The $I$-$V$ characteristics are calculated non-self-consistently by evaluating the transmission coefficient
over an effective bias-dependent Hamiltonian matrix, which in turns is obtained by adding a rigid shift to the zero bias Hamiltonian matrix 
elements of the electrodes and a linear potential across the insulating barrier. This is a good approximation to the self-consistent 
potential drop for tunnel junctions \cite{Ivan-2009}, which appears essentially like that of a standard parallel plate capacitor.

\begin{figure}
\includegraphics[width=0.45\textwidth,clip=true]{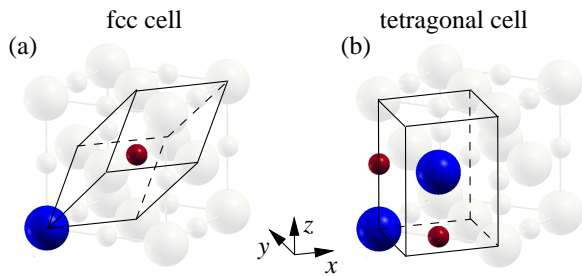}
\caption{(Color on line) Lattice structure of EuO constructed respectively with (a) the primitive fcc unit cell and (b) the tetragonal unit cell. 
The shaded grey atoms indicate the full cubic cell. Atom code: large spheres (blue)~=~Eu, small spheres~(red)=~O.}
\label{fig_struct}
\end{figure}

EuO crystallizes in the rocksalt structure with a lattice constant of $a_\mathrm{EuO}=5.144$ \AA. The primitive face-centered cubic 
(fcc) unit cell containing one Eu and one O atom is shown in Fig.~\ref{fig_struct}(a). Since we consider transport along the [001] direction, 
we have to use as basic building block for the EuO spacer a cell with size and orientation different from that of the primitive one. The 
smallest possible cell thus has tetragonal symmetry, and contains 2 Eu and 2 O atoms [see Fig.~\ref{fig_struct}(b)]. By stacking multiples
of such tetragonal cells along the $z$-direction we can construct barriers of arbitrary thickness, where one cell contains 2 EuO monolayers 
(MLs).

\begin{figure}
\includegraphics[width=0.45\textwidth,clip=true]{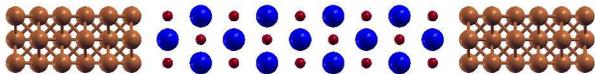}
\caption{(Color online) Supercell used for the spin filter tunnel junction. This consists of 6 Cu MLs (left lead), 9 Eu MLs (scattering region) and 5 Cu MLs 
(right lead). Atom code: large spheres (blue)~=~Eu, medium spheres (brown)~=~Cu, small spheres (red)~=~O.}
\label{fig_sftj_struct}
\end{figure}

The model spin filter junction considered here consists of $n$ MLs of EuO sandwiched between non-magnetic Cu electrodes, $n$ being 
an integer. As electrodes material we consider fcc Cu oriented along the [001] direction (lattice constant 3.61~\AA), a material choice which 
has been adopted in several experiments \cite{E.Negusse-J.Appl-2006,SANTOS-2008,Watson-2008,E.Negusse-J.Appl-2009}.
When the tetragonal EuO unit cell is used the lattice constant of Cu can be matched with only a slight strain to the EuO lattice, since the 
dimensions along $x$ and $y$ are $a_{\rm EuO}/\sqrt{2}=3.63$ \AA. The junction setup is illustrated in Fig.~\ref{fig_sftj_struct}. 
In this basic setup both the O and Eu atoms are placed over the hollow sites of the Cu surface. We also perform calculations for EuO shifted
in the $x$-$y$ plane. However the transport and spin filter properties are not sensitively dependent on such a shift. The equilibrium interface 
distance, $d$, between Cu and EuO is found to be $d$~=~2.8~\AA. Note that we do not consider possible oxidation at the Cu/EuO interface, 
or the formation of secondary EuO phases.  

\section{E\lowercase{u}O electronic structure}
\label{sec:bands}

\begin{figure}
\includegraphics[width=0.48\textwidth,clip=true]{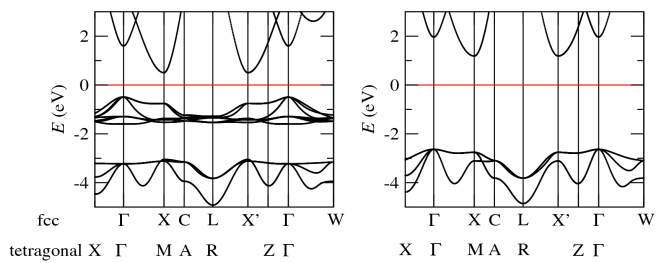}
\caption{LDA+U band structure of EuO for the primitive fcc unit cell [see Fig.\ \ref{fig_struct}(a)]: left panel majority spin bands; right panel minority 
spin bands. The band structure shows an indirect band gap from $\Gamma$ to X of about 1.0 eV.}
\label{fig_band}
\end{figure}

\begin{figure}
\includegraphics[width=0.48\textwidth,clip=true]{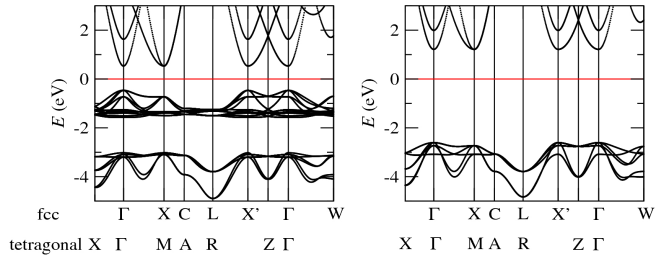}
\caption{LDA+U band structure of EuO for the primitive tetragonal unit cell [see Fig.\ \ref{fig_struct}(b)], with majority spin bands in the left panel and 
minority spin bands in the right panel. Due to the doubled size unit cell, as compared to the primitive fcc one, a band down-folding makes the $\Gamma$ 
point and the fcc X$^\prime$ point equivalent. This leads to a direct gap at $\Gamma$.}
\label{bands_EuO_transport_cell}
\end{figure}

Let us start our discussion by describing the electronic structure of EuO. The calculated EuO band structures for both the primitive 
fcc and the tetragonal unit cells are presented  respectively in Figs.~\ref{fig_band} and \ref{bands_EuO_transport_cell}. The results 
agree rather well with previously published calculations \cite{ingle08,D.B.Ghosh}. In order to compare the band structures for two 
cells we use the same $k$-space path. Standard high-symmetry points are specified for the primitive fcc BZ and also for the primitive
tetragonal BZ (where applicable). We note that due to band-folding for the tetragonal cell we find twice the number of bands than for 
the primitive fcc one. The X point [located at ($\pi$/a$_\mathrm{EuO}$,0,0)] and the X$^\prime$ point [located at 
(0,0,$\pi$/a$_\mathrm{EuO}$)] are equivalent in the fcc BZ. However for the tetragonal cell they are independent since the X$^\prime$ 
point is equivalent to $\Gamma$ in the tetragonal cell. There is a band gap of about 1.0~eV for majority spins (Fig.~\ref{fig_band}), 
whereas the gap increases to about 3.5 eV~for the minority ones. For the primitive fcc cell (Fig.~\ref{fig_band}) we find an indirect gap, 
where the top of the valence band is at $\Gamma$ and the bottom of the conduction band at X$^\prime$ point. This gap become direct
when the primitive tetragonal cell is used due to band-folding  (Fig.\ \ref{bands_EuO_transport_cell}). Note that the tetragonal cell is the
relevant one for the transport so that EuO along the [001] direction behaves as a direct gap ferromagnetic semiconductor. 

\begin{figure}
\includegraphics[width=0.3\textwidth,clip=true]{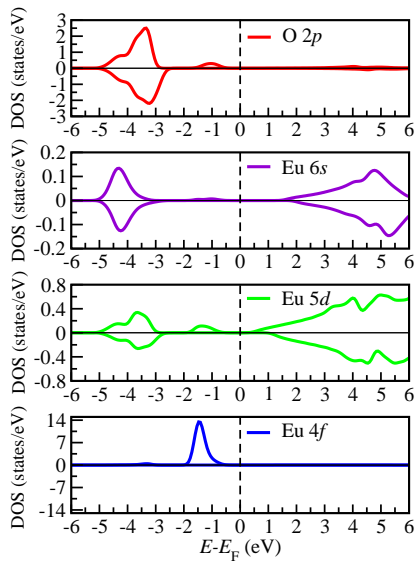}
\caption{(Color online) EuO density of states projected over the following atomic orbitals: O-2$p$, Eu-6$s$, Eu-5$d$ and Eu-4$f$.}
\label{fig_dos}
\end{figure}

The majority spins the energy spectrum consists of three parts: the conduction bands, the top valence bands, and the lower valence bands. 
The top valence bands are found in the range between $-1.6$ eV and $-0.5$ eV (we set the energy zero approximately to the middle of the gap). 
It is clearly seen from Fig.\ \ref{fig_dos}, where we show the projected DOS (PDOS) on different orbitals, that the main character of the top 
valence band is given by the Eu-4$f$ states. While many of the Eu-4$f$ bands show very little dispersion, indicating strong localization, 
in EuO the Eu-4$f$/O-2$p$ mixing leads to a significant dispersion around the $\Gamma$ point. This is an indication of delocalized states. 
We note that at the top of the valence band Eu-4$f$ and O-2$p$ contribute approximately equally to the density of states. In addition, the 
hybridization leads to a large contribution to the inter-site coupling of kinetic processes \cite{KUNES}. We therefore expect such delocalized states
to contribute significantly to the transport through a EuO barrier. Below the top valence bands, the bottom valence bands are observed in the
energy range from $-4.9$ eV to $-3.0$ eV. It is seen from Fig.\ \ref{fig_band} that the bottom valence bands are dominated by O-2$p$ states. 
The bandwidth of the O-2$p$ states (1.9 eV) is not much larger than the one of the Eu-4$f$ ones (1.1~eV), which indicates a similar wave-function 
delocalization. The energy gap between the top valence and bottom conduction bands is about 1.7 eV. The lower conduction bands are dominated 
by Eu-5$d$ states. Above 1.6~eV we find also contributions from the Eu-6$s$ orbitals. We note that the DOS at the lower end of the conduction bands
is very small, which is due to a high dispersion.


Next we consider the electronic band structure of the minority spins. The empty Eu-4$f$ states start at around 11.1~eV, so that they are not 
expected to affect the transport properties. As for the majority states, the conduction bands are dominated by Eu-5$d$ and Eu-6$s$ orbitals,
while the valence bands are dominated by O-2$p$. The spin filter character of EuO is due to a spin splitting of the bands, which leads to a 
difference in the band gaps for the majority and minority spins. The splitting between the majority and minority spin conduction bands is 
calculated to be about 0.6~eV, in agreement with previous theoretical studies \cite{ingle08,D.B.Ghosh,KUNES,Miyazaki-2008,Bousquet-2010,J.M.An-2011} 
and with recent experiments using 3D angle resolved photoemission spectroscopy \cite{Miyazaki-2009-Jpn,Miyazaki}. The Eu-6$s$ states show a similar spin splitting as the Eu-5$d$ states, namely 0.4~eV.

We now analyze the symmetries of the different bands. Symmetries are important for the transport through epitaxial tunnel junctions 
\cite{Butler-2001,N.M.Caffrey-2011}, since, at any given energy, only evanescent states in the insulating barrier with matching symmetry with 
propagating Bloch states in the electrodes can contribute to the conductance. Since the EuO gap in the tetragonal cell is at $\Gamma$, the transport 
will be dominated by states close to $k_x=k_y=0$, denoted as $\Gamma_{\mathrm{2D}}$, for which the barrier height is smallest (the coordinate 
system is defined in Fig.\ \ref{fig_struct}). In the energy range comprised between $-1.8$~eV and $1.6$~eV around $E_F$ the Cu electrodes have only
states of $\Delta_1$ symmetry with respect to the $z$-axis. For a cubic space group, the $\Delta_1$ symmetry transforms as a linear combination 
of 1 ($s$-orbitals), $z$ ($p_z$ orbitals), $2 z^2-x^2-y^2$ ($d_{z^2}$ orbitals), and  $z(2 z^2-3 x^2-3 y^2)$ ($f_{z^3}$ orbitals). Above 1.6~eV the 
Cu states have $\Delta_5$ symmetry. The $\Delta_5$ symmetry transforms as a linear combination of $x$ ($p_x$ orbitals), $y$ ($p_y$ orbitals), 
$xz$ ($d_{xz}$ orbitals) and $yz$ ($d_{yz}$ orbitals). At energies below $-1.8$ eV we find the Cu-3$d$ orbitals, so that there are also states
with different symmetries. For a transport measurement up to about 2~V we therefore expect the states with $\Delta_1$ symmetry to determine 
the transport properties.

\begin{figure}
\includegraphics[width=0.45\textwidth,clip=true]{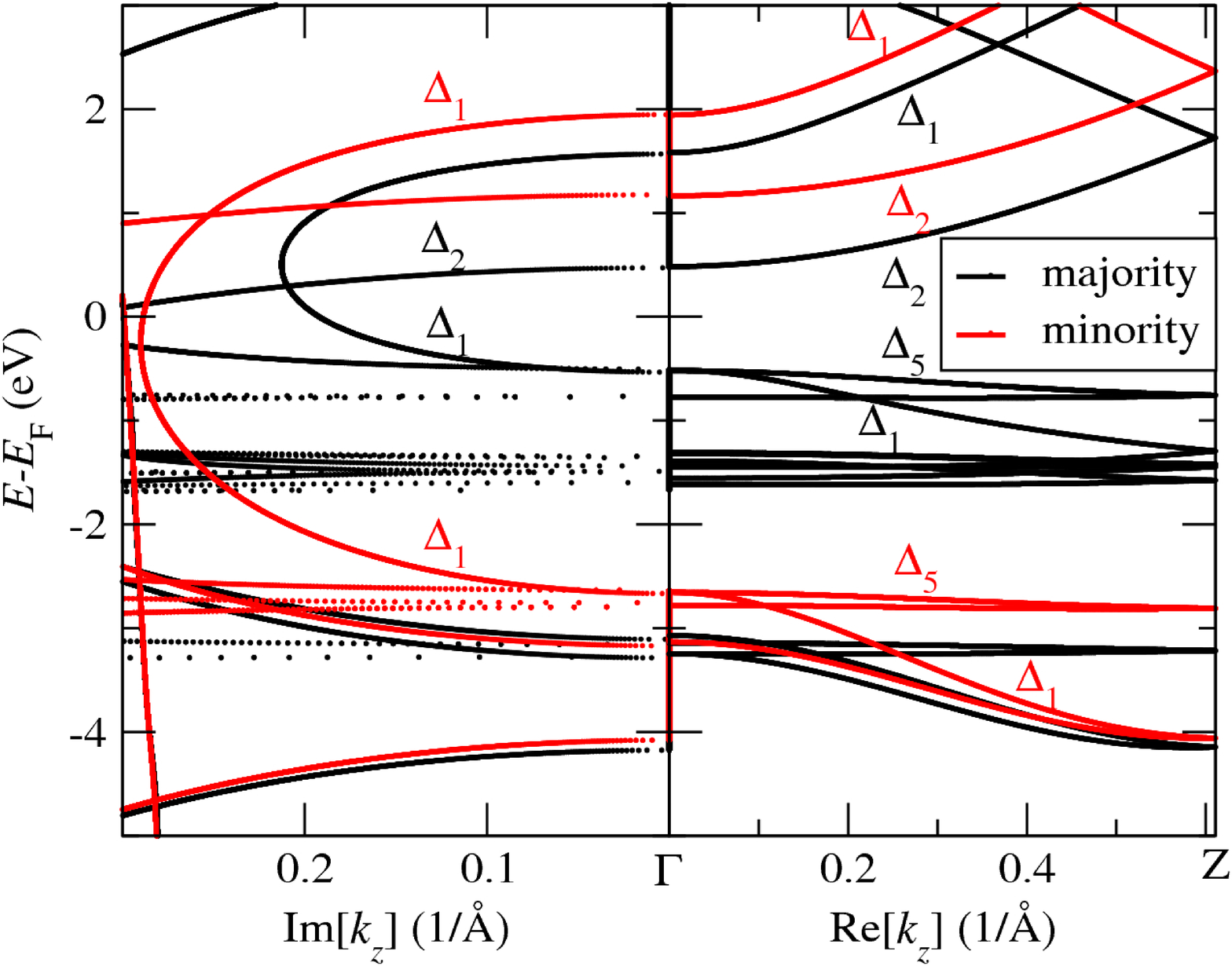}
\caption{(Color online) Complex band structure of EuO in its tetragonal cell [see Fig.\ \ref{fig_struct}(a), where the symmetries are indicated for the bands 
relevant to our transport setup. Red lines~=~minority spins, black lines~=~majority spins.}
\label{fig_cbs}
\end{figure}

In order to investigate in more details the propagating and evanescent states in the EuO barrier we calculate the complex band structure (Fig.\ \ref{fig_cbs}) 
at $\Gamma_{\mathrm{2D}}$ \cite{Butler-2001,N.M.Caffrey-2011,ddtcormac} and analyze the symmetries of the evanescent states. Real wave vectors 
($\mathrm{Im}[k_z]=0$) represent propagating states and complex wave vectors ($\mathrm{Im}[k_z]\neq0$ represent evanescent states, since their 
wave functions decay as $\mathrm{exp}(-\mathrm{Im}[k_z]\cdot{z})$ across the barrier. For the majority spins the top of the valence band is three-fold 
degenerate, with one state with $\Delta_1$ symmetry (Eu-4$f_{z^3}$ and O-2$p_z$ orbitals) and two states with $\Delta_5$ symmetry. Since the 
state with $\Delta_1$ symmetry has a lower effective mass than those with $\Delta_5$ symmetry, the corresponding $\Delta_1$ evanescent states 
have smaller $\mathrm{Im}[k_z]$, and therefore a slower decay. The bottom of the conduction band is given by a state with $\Delta_2$ symmetry
(Eu-5$d_{x^2-y^2}$ orbital). At 1.6~eV we find the Eu-4$s$ states, which have $\Delta_1$ symmetry. Therefore $\mathrm{Im}[k_z]$ for the $\Delta_1$ 
band forms a semi-circle between the Eu-4$f_{z^3}$ states at the top of the valence band and the Eu-4$s$ states at 1.6~eV. 

Consequently, we expect the
tunneling transmission to be dominated by the $\Delta_1$ states in this energy range. The Eu-5$d$ propagating states with $\Delta_2$ symmetry cannot 
couple to the $\Delta_1$ Cu states, and therefore are not expected to contribute significantly to the transmission. For $k_x$ and $k_y$ different from 
zero the Eu-5$d$ states are no longer fully orthogonal to the Cu-4$s$ states. However, since the gap increases with increasing $k_x$ and $k_y$, the 
barrier height for such states is larger. Hence their contributions to the total transmission are smaller. Overall one expects a rather weak coupling to the 
Cu-4$s$ states and a low transmission for states at the bottom of the EuO conduction band.

In the minority spins bands there are also three degenerate states at the top of the valence band, one with $\Delta_1$ symmetry (O-2$p_z$ orbital) and 
two with $\Delta_5$ symmetry (O-2$p_x$ and O-2$p_y$ orbitals). The $\Delta_1$ state again has lower effective mass and therefore a smaller decay for 
evanescent states. At the bottom of the conduction band the symmetries are analogous to those of the majority spins. We therefore also expect the 
minority conductance to be dominated by the $\Delta_1$ symmetry. However, since the minority band gap is much larger than that of the majority 
spins, $\mathrm{Im}[k_z]$ is also significantly larger for $\Delta_1$ states in the gap. As a consequence, we expect $T^\uparrow$ to be substantially larger 
than $T^\downarrow$. This is usually attributed only to different energies of the bottom of the conduction band \cite{J.S.Moodera-2007}, whereas our complex 
band structure analysis indicates that the decay of the evanescent states has significant contributions also from states in the valence band, especially for the 
majority spins. We note that the Eu-5$d$ $\Delta_2$ states will contribute to the transport if different electrodes, possessing $\Delta_2$ symmetry states in 
the relevant energy range, are used

\section{Spin transport properties of the E\lowercase{u}O junction at zero bias}
\label{sec:T_0}

\begin{figure}
\includegraphics[width=0.45\textwidth,clip=true]{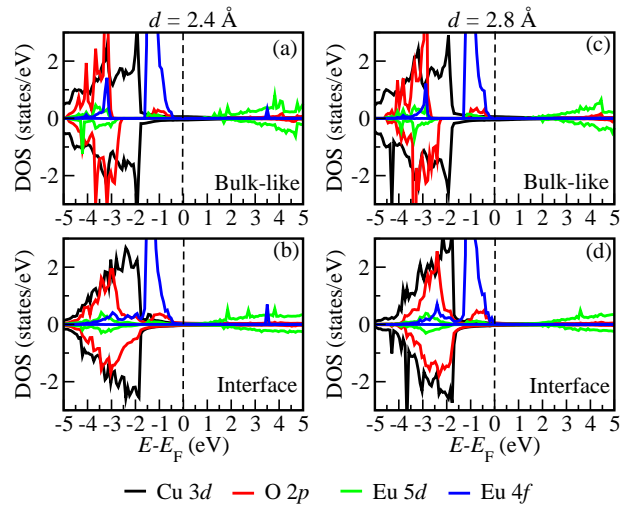}
\caption{(Color online) PDOS for Cu, O and Eu for a junction comprising 9 MLs EuO. The PDOS is calculated at the center of the junction (a) and 
the interface (b) for $d=2.4$~\AA, and at the center of the junction (c) and the interface (d) for $d=2.8$~\AA. Here $d$ is the distance between the
Cu and the EuO planes at the EuO/Cu interface.}   
\label{Total-PDOS}
\end{figure}

In this section we analyze the zero bias transport properties of Cu/EuO/Cu junctions by taking as an example a stack containing 9 MLs of 
EuO ($t=20.58$~\AA). The energy level alignment between the metal and the insulator is an important factor determining the transport 
properties of the junction. To a first approximation, one can estimate the alignment by comparing the workfunctions, $W$, of the two components. 
The workfunctions of EuO and Cu are calculated by using the Hartree electrostatic potential, $V_H$, as the reference potential \cite{Ivan-2010}. 
We define $W_\mathrm{Cu}$ as the difference between $E_F$ and the vacuum potential $V_\mathrm{vacuum}$ of a Cu slab, while $W_\mathrm{EuO}$ 
is given by the difference between $V_\mathrm{vacuum}$ and the energy of the valence band top of a EuO slab, $V_\mathrm{VB}$. We calculate 
$W_\mathrm{Cu}=3.9$~eV and $W_\mathrm{EuO}=1.8$~eV, both values in good agreement with the experimental values, respectively ranging between 
4.5~eV to 5.1~eV for Cu, and being 1.7~eV for EuO \cite{D.E. Eastman}. Since $W_\mathrm{EuO}<W_\mathrm{Cu}$, electron transfer from EuO to Cu 
can be expected at the interface, leading to the pinning of the valence band top of EuO to the Cu $E_F$. This is indeed the case for the equilibrium distance 
($d=2.8$ \AA), where the Eu-4$f$ states are located just below $E_F$ [see Fig.\ \ref{Total-PDOS}(c) and (d)].
 
In experiments investigating bias-dependent transport the actual junctions usually contain polycrystalline EuO, so that the interface between Cu and 
EuO is not well defined \cite{Martina-Europhys-2009}. In our calculations, in contrast, we assume a perfect epitaxial interface and do not explicitly 
consider the formation of defects, oxidation of Cu, or the possible formation of Eu$_2$O$_3$ at the interface \cite{E.Negusse-J.Appl-2006,Martina-J.Appl-2009}. 
The effect of such modifications of the interface on the electronic structure are manyfold, the most important being that they usually lead to a different 
charge transfer and therefore to a different band alignment between Cu and EuO. In practice this means that (depending on the detailed structure of the 
interface, which is determined by the experimental conditions) $E_F$ can be placed at different positions across the EuO gap. In our calculations we can
tune the charge transfer and consequently the position of $E_F$ in the EuO gap by modifying the distance, $d$, between Cu and EuO at the interface. We 
find that decreasing the distance from the equilibrium one, i.e., increasing the coupling between Cu and EuO, the EuO states shift to lower energies with 
respect to $E_F$. For $d=2.4$ \AA\ $E_F$ is located approximately in the middle of the gap [see Fig.\ \ref{Total-PDOS}(a) and (b)], while for $d=2.2$ \AA\ it 
is pinned at the EuO conduction band minimum (not shown). Therefore, $d=2.2$ \AA\ can be used to simulate the transport properties of n-type EuO, as 
obtained for O-deficient barriers \cite{M.Barbagallo,X.Wang,J.A.C.Santana}. In the remaining part of this paper we will present the transport properties for both $d=2.4$ \AA\ and $d=2.8$ \AA, in order to illustrate the effect of a shift of $E_F$ induced by interface modifications. Importantly, we will demonstrate 
that, for any position of $E_F$, EuO always shows excellent spin filter characteristics up to high bias voltages. We note that if experimentally a perfect 
epitaxial junction can be realized, the measurements should correspond to our results for the equilibrium distance $d=2.8$ \AA.

In Fig.\ \ref{Total-PDOS} we compare the PDOS of atoms at the Cu/EuO interface with the PDOS of atoms at a maximal distance from the interface 
(to be considered bulk-like). While the two PDOSs are similar for most atomic orbitals, there is a significant difference for the O-2$p$ states. At the 
interface the O-2$p$ states extend over a larger energy range as compared to bulk-like atoms, indicating a coupling to the Cu substrate. In fact, for 
$d=2.4$~\AA\ the broadening is more pronounced than for $d=2.8$~\AA, which is due to the larger coupling. The PDOS in the middle of the EuO layer
is very similar to that of bulk EuO, indicating that in the middle of the junction one recovers the bulk electronic structure of EuO. Although in 
Fig.\ \ref{Total-PDOS} only the fully occupied Cu-3$d$ states are shown, we note that the Cu-4$s$ states determine the transport properties of the 
electrodes, since they have an approximately constant PDOS in a large energy range around $E_F$. In Fig.\ \ref{Total-PDOS} the Cu-4$s$ PDOS is 
not shown, since it is not visible on the chosen scale.
\begin{figure}
\includegraphics[width=0.475\textwidth,clip=true]{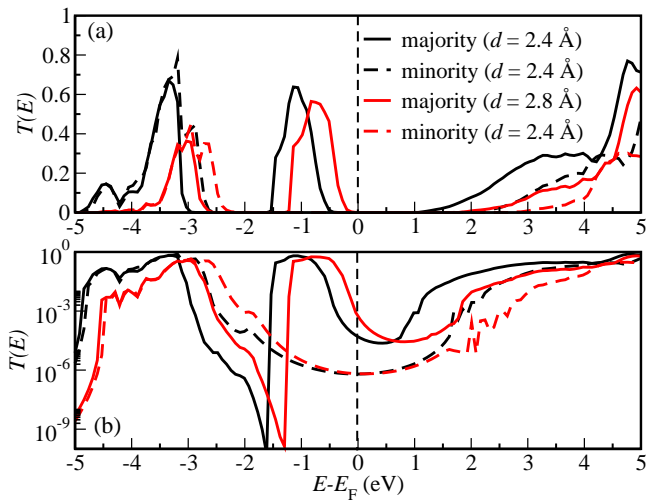}
\caption{Transmission coefficient of a Cu/EuO/Cu junction formed by 9 EuO MLs for $d=2.4$~\AA\ and $d=2.8$~\AA, plotted on a linear (a) and on 
a logarithmic scale (b). Note that for $d=2.8$ \AA, $T(E)$ shifts to higher energy when compared to the case of $d=2.4$ \AA.}
\label{Compare-9EuO-d24-d28}
\end{figure}

The zero bias transmission coefficient $T(E)$ for different $d$ is shown in Fig.~\ref{Compare-9EuO-d24-d28} [on a linear scale in panel (a) and on a 
logarithmic scale in panel (b)]. Due to the shift of the electronic states for different $d$ (see Fig.\ \ref{Total-PDOS}), $T(E)$ is shifted towards higher 
energies for $d=2.8$ \AA\ as compared to $d=2.4$ \AA, while there is only a minor change in the height of the different transmission peaks. It can be 
seen that $T(E)$ has two small gaps for the majority spins and one large gap for the the minority ones. For the minority spins the conduction occurs 
only through the O-2$p$ states for energies below $-2.5$~eV and through the Eu-5$d$ and Eu-6$s$ states for energies above about 2~eV. In 
contrast, for the majority spins the Eu-4$f$ states just below $E_F$ also contribute significantly to the conductance. This is consistent with the band 
structure, which shows that the Eu-4$f$ states hybridize with the O-2$p$ and are rather delocalized. The result of such hybridization is a large majority 
transmission in the energy range between $-1.5$~eV and $E_F$. Therefore, for all energies below $E_F$ (down to about $-1.5$~eV) the transmission 
of the majority spins is much larger than that of the minority ones. For energies above $E_F$ (up to about 2~eV) it is also significantly larger. This is 
due to fact that the Eu-5$d$ conduction band minimum is located about 0.6~eV lower in energy for the majority spins than for the minority ones 
(Fig.\ \ref{fig_cbs}). 

The overall result hence is that for any position of $E_F$ in the EuO gap we expect a very high spin-polarization of the current. This should persist up to 
high bias voltages (of the order of the energy for which the transmission of the majority spins is much larger than that of the minority ones). The spin-filtering
efficiency at small applied bias is defined as $[T^{\uparrow}(E_F)- T^{\downarrow}(E_F)]/T(E_F)$ \cite{Shen}. Since $T^\uparrow(E_F)$ is about two 
to three orders of magnitude larger than $T^\downarrow(E_F)$, we have a spin filter efficiency close to 100\%, indicating that the EuO barrier is an 
almost perfect spin filter. We note that the extremely high efficiency is obtained for defect-free epitaxial junctions. Such value might be reduced by 
defects in the EuO barrier as well as for polycrystalline EuO. Nevertheless one can expect that also in these cases the spin-filter efficiency may remain 
high.

\begin{figure}
\includegraphics[width=0.475\textwidth,clip=true]{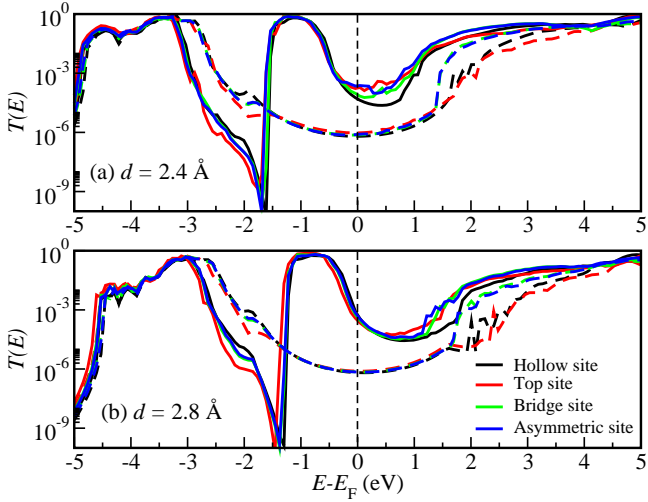}
\caption{(Color online) Transmission coefficient for a Cu/EuO/Cu junction with a 9 MLs EuO barrier and with different interface geometries: 
(a) $d=2.4$ \AA\ and (b) $d=2.8$ \AA. The solid lines represent the majority spin transmission, whereas the dashed ones are for minority 
spins. Note that overall the transmission depends little on the lateral position of the O atoms with respect to the Cu surface.}
\label{TRC-d24-28-9EuO-Hallow-Top-Bridg-Asym}
\end{figure}

In order to investigate the dependence of the spin transport properties on the geometry of the Cu/EuO interface we calculate $T(E)$ for different 
positions of EuO with respect to Cu. So far we have considered a setup in which O is placed on top of the hollow site of the Cu surface. We compare the results with the following three geometries: 1) O placed on top of Cu, 2) O placed on the bridge site between two Cu atoms, and 
3) O placed at an arbitrary low symmetry site. The equilibrium distance is 2.8 \AA\ for all cases and the total energies are similar, with the bridge site 
having the lowest energy. In Fig.\ \ref{TRC-d24-28-9EuO-Hallow-Top-Bridg-Asym} we show $T(E)$ calculated for the different sites, for both 
$d=2.4$ \AA\ and $d=2.8$ \AA. Overall the changes in transmission corresponding to the different sites are rather small, a fact, which indicates 
that the geometry of the Cu/EuO surface does not significantly affect the band alignment and transport properties of the junction.

At the bottom of the conduction band there are some quantitative differences in transmission between the different sites. This is due to the fact 
that shifting the O atom with respect to Cu alters the coupling between the Cu-4$s$ and Eu-5$d$ states. As discussed in Section\ \ref{sec:bands},
while the states on the top of the valence band couple well to the $\Delta_1$ Cu band, the states at the bottom of the conduction band couple only 
slightly to these states due to a symmetry mismatch. This is also the reason for the fact that the transmission gap is somewhat larger than the 
EuO band gap, and for the slow increase of the transmission at the bottom of the conduction band with energy. For the bridge and low symmetry 
sites the symmetry mismatch is slightly reduced, resulting in a somewhat smaller transmission gap. Since the results for the different sites are 
very similar, we will only consider the hollow site in the remaining of the paper.

\section{Thickness dependence of the conductance}

\begin{figure}
\includegraphics[width=0.475\textwidth,clip=true]{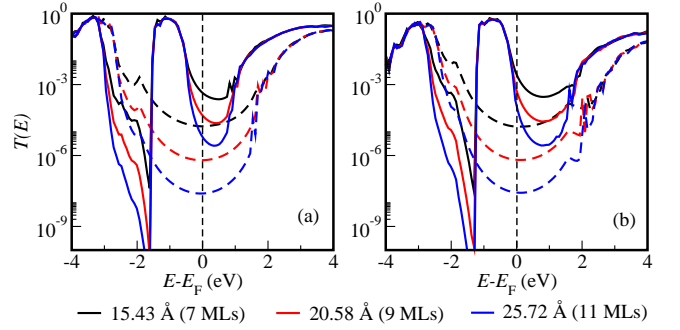}
\caption{(Color online) Thickness dependence of $T(E)$ at zero bias for (a) $d=2.4$~\AA\ and (b) $d=2.8$~\AA. The solid and dashed lines refer to the 
majority and minority spins, respectively. At $d=2.4$~\AA\ the tunneling gap of the majority spins is found from $-0.45$~eV to 1.1~eV, while for the 
minority ones it extends from $-2.5$~eV to 1.6~eV. At $d=2.8$~\AA\ the tunneling gaps of the majority and minority spins for all thicknesses range 
between $-0.1$~eV to 1.75~eV and $-2.3$~eV to 2.5~eV, respectively.}
\label{TRC-d24-d28-5eV} 
\end{figure}

In this section we analyze the effect of the EuO thickness, $t$, on the spin transport properties of the Cu/EuO/Cu junction. The transmission spectra 
for different EuO thicknesses, and for both $d=2.4$ \AA\ and $d=2.8$ \AA, are shown in Fig.\ \ref{TRC-d24-d28-5eV}. From the figure it can be seen 
that the band alignment is not affected by the thickness, while the transmission in the gap decreases exponentially with it. Since for energies in the 
gap we have $T^\sigma(E)\propto \exp[-2\kappa^\sigma(E) t]$, we can estimate $\kappa^\sigma$ from the change of $T^\sigma(E)$ with increasing $t$. 
Indeed $\kappa^{\sigma}$ can be calculated from $T^{\sigma}$ evaluated at two different thicknesses $t_1$ and $t_2$ as 
\begin{equation}
\kappa^\sigma(E) = \frac{1}{2(t_2-t_1)}\mathrm{ln}\left[\frac{T_1^\sigma(E)}{T_2^\sigma(E)}\right],
\label{eq:kappa}
\end{equation}
where $T_1^{\sigma}$ and $T_2^{\sigma}$ are the transmissions for $t_1$ and $t_2$, respectively.
\begin{figure}
\includegraphics[width=0.475\textwidth,clip=true]{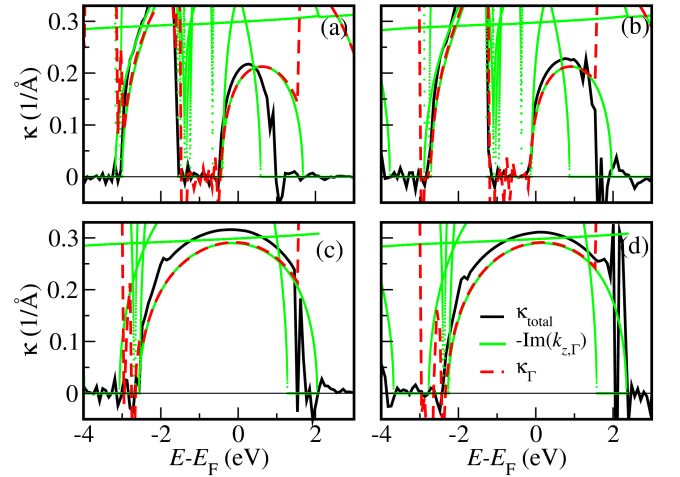}
\caption{Damping coefficient $\kappa$ calculated from $T(E)$ for the majority [(a) and (b)] and minority spins [(c) and (d)] at $d=2.4$~\AA\ and 
$d=2.8$~\AA. $\kappa$ is calculated from Eq.~\ref{eq:kappa} by using the transmission coefficient calculated for 9 and 11 EuO MLs. The solid 
lines refer to the total transmissions and the dashed lines to the transmissions at $k_x=k_y=0$ only. The green lines show $\mathrm{Im}[k_z]$ 
from Fig.\ \ref{fig_cbs}, which is also calculated for $k_x=k_y=0$.}
\label{Damping-Coeff-Up-Dn-Total}
\end{figure} 

In Fig.\ \ref{Damping-Coeff-Up-Dn-Total} we show the calculated $\kappa(E)$ (solid lines), evaluated for $t_1=20.58$~\AA\ (9 EuO MLs) and 
$t_2=25.72$~\AA\ (11 EuO MLs). For energies within $\pm$~1.5~eV around $E_F$, $\kappa^\uparrow$ is significantly smaller than $\kappa^\downarrow$. 
Therefore, increasing the EuO layer thickness leads to an enhancement of the ratio between the transmissions of the majority and minority spins, i.e., to an 
increase of the spin filtering efficiency. For $d=2.4$~\AA, $\kappa^\uparrow$ is larger than zero in the range from $-0.5$~eV to 1.0~eV (resulting in a 
transmission gap of 1.5~eV), whereas $\kappa^\downarrow$ is larger than zero in the range from $-2.6$~eV to 1.6~eV (resulting in a transmission 
gap of 4.2~eV). For $d=2.8$~\AA, $\kappa$ is shifted to higher energies, and the transmission gap amounts to about 1.7~eV for the majority spins
and 4.7~eV for the minority ones. While $\kappa$ shows a parabolic behavior for energies close to the valence band top, for the conduction band 
minimum the behavior is less well defined. This is due to the symmetry mismatch between the Cu $\Delta_1$ states and the conduction band Eu-5$d$ 
states, see Sec.\ \ref{sec:T_0}. Nevertheless the barrier height is about 0.6~eV larger for the minority spins as compared to the majority ones for 
transmission through the conduction band. For the transmission through the valence band the difference is even larger, since there are no filled 
minority spin Eu-4$f$ states.

In Fig.\ \ref{Damping-Coeff-Up-Dn-Total} we also present the $\kappa(E)$ obtained from the transmission only at the $\Gamma_\mathrm{2D}$ point 
(dashed curve). Close to the valence band the wave-function decay is similar to the previous case, which shows that $T(E)$ is dominated by 
contributions around $\Gamma_\mathrm{2D}$. However, for energies close to the conduction band, $\kappa(E)$ at $\Gamma_\mathrm{2D}$ is
much larger than the total $\kappa(E)$, which shows that here the transport occurs mainly at $k$-points away from the BZ center. A comparison 
with the EuO complex bands (green curves; see Fig.\ \ref{fig_cbs} for a description of the symmetries) shows that, at $\Gamma_\mathrm{2D}$, 
$\kappa(E)$ follows approximately $\mathrm{Im}[k_z]$ for $\Delta_1$ states. It is clear that no transmission occurs through the conduction 
band $\Delta_2$ states. For $k$-points away from $\Gamma_\mathrm{2D}$ the EuO $\Delta_2$ states can couple to the Cu states, which leads 
to a decrease of the total $\kappa(E)$ above the conduction band minimum. For electrode materials with $\Delta_2$ states above $E_F$ we expect 
the total $\kappa(E)$ to follow the $\Delta_2$ complex bands at $\Gamma_\mathrm{2D}$ for energies below the conduction band minimum.

The barrier heights are calculated from the damping coefficient shown in Fig.\ \ref{Damping-Coeff-Up-Dn-Total}. The values of  $\Phi^\uparrow_\mathrm{CB}$ and  $\Phi^\downarrow_\mathrm{CB}$ are determined as the lowest energy at which $\kappa(E)$ crosses zero in the conduction band for the majority and minority spins, respectively. Similarly, $\Phi^\uparrow_\mathrm{VB}$ and $\Phi^\downarrow_\mathrm{VB}$ are determined as the highest energy at which $\kappa(E)$ crosses zero in valence band for the majority and minority spins, respectively. By this method we obtain for $d=2.4$~\AA\ the values $\Phi^\uparrow_\mathrm{CB}=1.0$ eV, $\Phi^\downarrow_\mathrm{CB}=1.6$ eV, $\Phi^\uparrow_\mathrm{VB}=0.5$ eV, and $\Phi^\downarrow_\mathrm{VB}=2.6$ eV. For $d=2.8$~\AA\ we obtain $\Phi^\uparrow_\mathrm{CB}=1.6$ eV, $\Phi^\downarrow_\mathrm{CB}=2.2$ eV, $\Phi^\uparrow_\mathrm{VB}=0.1$ eV, and $\Phi^\downarrow_\mathrm{VB}=2.4$ eV. The exchange splitting for the conduction band of 0.6 eV matches the value for bulk EuO. 
%
%
\begin{figure}
\includegraphics[width=0.475\textwidth,clip=true]{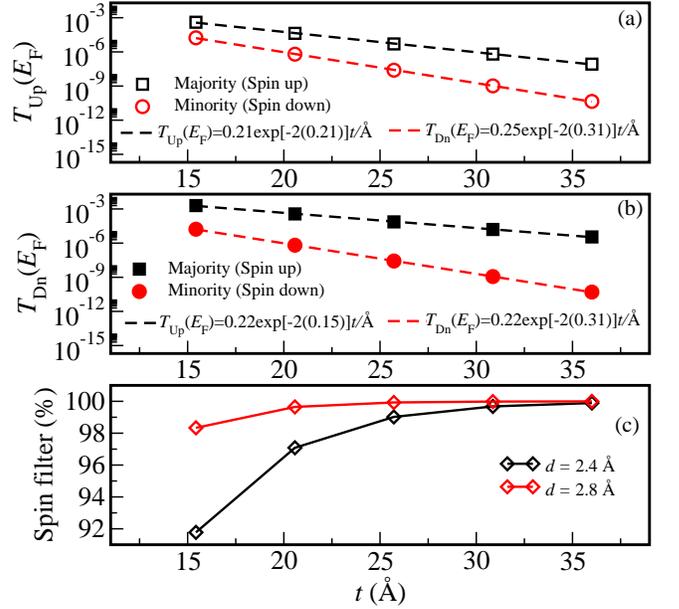}
\caption{Dependence of $T(E_F)$ on $t$ for (a) $d=2.4$~\AA\ and (b) $d=2.8$~\AA. The spin filtering efficiency at $E_F$ is shown in (c) for both values of $d$.}
\label{Joint-TRC-Effective-at-Fermi}
\end{figure}  

For the low bias conductance the thickness dependence of the transmission coefficient at $E_F$ is evaluated and from it the decay coefficient. In Fig.~\ref{Joint-TRC-Effective-at-Fermi} we show $T(E_F)$ at $d=2.4$~\AA\ and $d=2.8$~\AA\ 
as functions of the EuO thickness. The decay rates correspond to those obtained in Fig.\ \ref{Damping-Coeff-Up-Dn-Total} at $E_F$. In Fig.\ \ref{Joint-TRC-Effective-at-Fermi}(c) the effective spin filtering at $E_F$ is addressed. It can be seen that it increases towards 100\% as $t$ increases.           

\section{The $I$-$V$ curve of the E\lowercase{u}O spin filter}

\begin{figure}
\includegraphics[width=0.475\textwidth,clip=true]{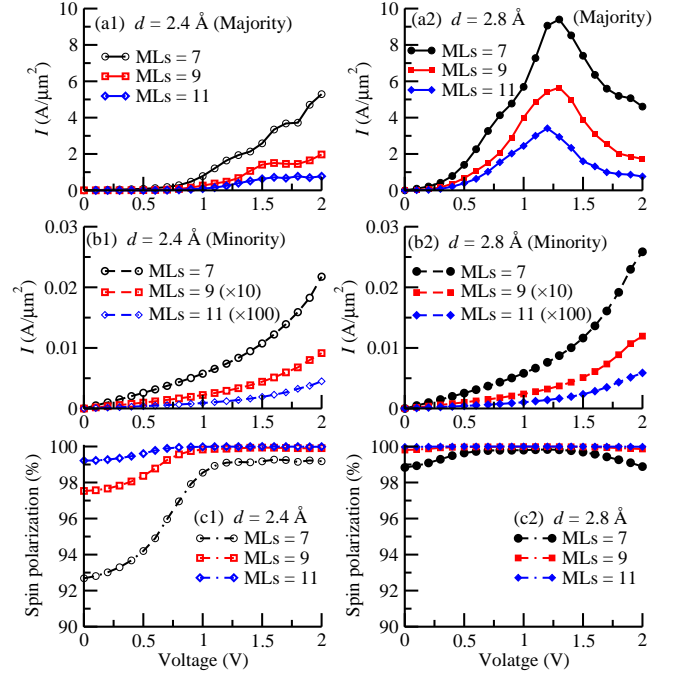}
\caption{Spin-polarized current and spin polarization as functions of the bias voltage
for different EuO thicknesses. (a1) Majority spin current at $d=2.4$~\AA, (a2) majority spin
current at $d=2.8$~\AA, (b1) minority spin current at $d=2.4$~\AA, (b2) minority spin
current at $d=2.8$~\AA, (c1) spin polarization at $d=2.4$~\AA, and (c2) spin polarization
at $d=2.8$~\AA.} 
\label{I-V-d24-d28-New}
\end{figure}
    
\begin{figure}
\includegraphics[width=0.475\textwidth,clip=true]{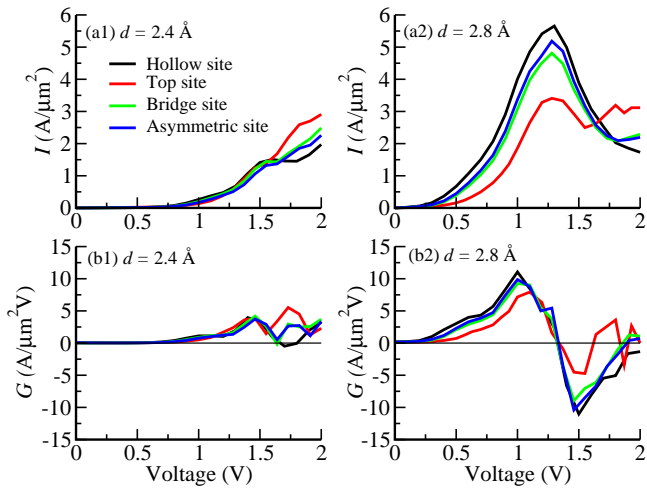}
\caption{Total current and conductance for a Cu/EuO/Cu junction with a 9 EuO MLs barrier at $d=2.4$ and $2.8$~\AA\ with
different interface geometries. (a1) Current at $d=2.4$~\AA, (a2) current at $d=2.8$~\AA, (b1) conductance at $2.4$~\AA, and 
(b2) conductance at $2.8$~\AA.}
\label{Current-dI-dV-EuO-Geometry}
\end{figure}

The spin-resolved current-voltage, $I$-$V$, characteristic and the spin polarization as functions of the bias voltage are shown in Fig.\ \ref{I-V-d24-d28-New} 
for 7, 9 and 11 MLs of EuO, and for both $d=2.4$ and $2.8$~\AA. The majority spin current is higher than the minority spin current for all bias voltages 
and for all thicknesses, leading to large polarization in all cases. At low bias there is a rapid increase of the tunneling current with the bias. This is typical
for such systems due to the fact that the current is not only determined by electrons at $E_F$ but also by those in the energy range $E_F \pm eV/2$ 
(bias window), for which the barrier height can be reduced up to about $eV/2$. At a bias of about 0.7~V for $d=2.4$~\AA\ the electrons start to flow 
through the valence band top, so that there is a sharp increase in current. Since for $d=2.8$~\AA\ the valence band is very close to $E_F$, the onset is 
found at very low bias. The current increases up to about 1.3~V, above which a decrease with increasing bias is found. This is caused by a reduction 
of the transmission over the entire bias window once the bias is very large. Such a current reduction results from the enhanced scattering as the potential 
is tilted inside the barrier due to the applied bias voltage. In these large scattering conditions, incoherent contributions to the current (not included here) are 
expected to play an important role.

For the minority spins, in contrast, the current remains in the tunneling regime for all bias voltages due to the large barrier height. The large difference between
majority and minority spin currents is reflected by a high spin polarization, $P$, 
defined as $\left(I^{\uparrow}-I^{\downarrow}\right)/\left(I^{\uparrow}+I^{\downarrow}\right)$ [see Fig.\ \ref{I-V-d24-d28-New}(c1)]. $P$ remains large for all 
the bias voltages considered due to the small $I^{\downarrow}$. We note that the spin splitting of the conduction band plays only a secondary role in our 
results, since the majority current is mainly determined by the valence band contribution (which is not sensitive to the exact position of $E_F$ inside the 
EuO energy gap). For n-type EuO one expects a pinning of the conduction band to $E_F$, leading to an almost metallic-like character dominated by the 
EuO conduction electrons.

The dependence of the $I$-$V$ characteristics and of the conductance trace $G(V)=dI/dV$ on the Cu/EuO interface geometry is addressed in 
Fig.\ \ref{Current-dI-dV-EuO-Geometry}. Overall the $I$-$V$ characteristics are similar for all the interfaces, which reflects the fact that the
zero bias transmission is also rather similar (see Fig.\ \ref{TRC-d24-28-9EuO-Hallow-Top-Bridg-Asym}). For $d=2.4$~\AA\ there is a sharp increase 
in $G$ at the current onset, whereas it then stays approximately constant. For $d=2.8$~\AA, in contrast, $G$ is large at low bias and changes sign at 1.3~V.

\section{Conclusions}
We have studied the electronic structure, the complex band structure and the spin transport properties of epitaxial Cu/EuO/Cu tunnel junctions. 
The spin transport properties of EuO epitaxially grown on Cu are dominated by the Eu-4$f$ valence states and by the Eu-5$d$ conduction states.
We show that EuO acts as an almost perfect spin filter, where close to 100\% spin polarization can be achieved. The polarization increases with 
increasing EuO thickness as expected from the complex band structure, where the decay of the wave function into EuO is predicted to be much 
smaller for the majority spins than for the minority ones. Since the conduction states of EuO have no $\Delta_1$ symmetry to match the states in 
the Cu electrodes, the gap in the transmission coefficient is significantly larger than the band gap of EuO. Under a bias voltage the spin polarization 
of the current does not decrease, as in usual tunnel junctions, but remains approximately constant up to all considered bias voltages. This opens 
promising perspectives of using EuO for device applications. 
 
\section{Acknowledgements}
We would like to acknowledge technical assistance from Apirat Siritarathiwat and Hao Wang. N.\ J.\ and U.\ E.\ acknowledges financial support by the Deutsche Forschungsgemeinschaft through TRR 80. Computational resources have been provided by TCHPC Ireland and LRZ Munich, Germany. I.\ R.\ and S.\ S.\ acknowledge
financial support by KAUST (ACRAB project).

\end{document}